\documentstyle[12pt]{article}

 \newcommand{\G}{{\cal G\mit}}
 \newtheorem{th}{Theorem}[section]
 \newtheorem{defin}[th]{Definition}
 \newtheorem{prop}[th]{Proposition}
 \newtheorem{lemma}[th]{Lemma}

\begin{document}

 \begin{titlepage}
   \title{ { \bf  Weinstein -- Xu  invariants and link group
           representations.  I  }       }

   \author{ \\ {\bf
     A. A. Balinsky} \thanks{e-mail: balin@leeor.technion.ac.il }  \\ \\
         Technion-Israel Institute of Technology \\
         Department of Mathematics \\
         32000 Haifa, ISRAEL}

     \date{ }

     \maketitle

     \thispagestyle{empty}

          \begin{abstract}
Topological interpretation of the link invariants
associated with the Weinstein--Xu classical solutions
of the quantum Yang-Baxter equation are provided.

          \end{abstract}

      \end{titlepage}


  \section*{Introduction.}


Our main goal in this paper is to give topological interpretation
for the Weinstein-Xu link invariants \cite{WX} in the case of
factorizable Poisson groups. The general case of an arbitrary
quasitriangular Poisson Lie group and its connection with Joyce's
theory of knot quandles will be subject of the next paper.

  In the last several years many approaches to the Jones
  polynomial and its generalizations have emerged. The article
  \cite{JB} give an excellent account of main developments in
  knot theory which followed upon the discovery of the Jones
  polynomials \cite{VJ} in 1984. But from our point of view, we
  nevertheless are far from the understanding of the topological
  meaning of the new invariants. It will be useful to recall here
  the following two unsolved problems \cite{Knot}

{\bf Problem 1. (J. Birman)}  {\em Does the Jones action have
geometric or algebraic meaning, as does the classical action } ?

{\bf Problem 2. (V. Jones)} {\em Give a natural way to derive the
Hecke algebra representation of the braid group from the usual
action of the braid group on the free group.}

   In a very deep paper \cite{WX} A. Weinstein and P. Xu defined a
   broad class of knot and links invariants using a kind of the
   classical solutions of the quantum Yang--Baxter equation. In
   \cite{WX} the classical analogue is developed for part of the
   standard construction in which generalized Jones invariants
   are produced from representations of quantum groups. As far as
   I know their article is the first attempt to  understand
   the topological meaning of the {\em quantum group
   invariants} on the quasi-classical level.

In this article we shall show that the Weinstein--Xu link
invariants in the case of factorizable Poisson Lie group coincide
with the space of representations of the link group into this
group. The restriction to a symplectic realization leads to the
space of representations for which the images of the meridians
belong to the corresponding conjugancy class .

This paper is organized as follows.

   In Sec. 1 we recall some properties of Poisson Lie groups and
   the Weinstein--Xu link invariants. For more details on this
   subject, see \cite{WX}.

   In Sec. 2 we describe the relation between fixed point sets of
   the braid groups action and representations of the link
   groups.

   In Sec. 3 it is proved that the Weinstein--Xu link invariants
   in the case of factorizable Poisson Lie group coincide with
   the space of representations of the link group into this
   group.

   In Sec. 4 some discussions are made.

\pagebreak

 \section{Weinstein--Xu invariant}

Let $G$ be a Poisson Lie group. This means that $G$ is a Lie
group equipped with a Poisson structure $\pi$ such that the
multiplication in $G$ viewed as map $G \times G \rightarrow G $
is a Poisson mapping, where $G \times G $ carries the product
Poisson structure. The theory of Poisson Lie groups is a
quasiclassical version of the theory of quantum groups.

  One can easily check that the Poisson structure $\pi$ must
  vanish at the identity $ e \in G$, so that its linearization
  $d_{e} \pi : \G \rightarrow \G \wedge \G$ at $e$ is well
  defined ( here $\G$ is the Lie algebra of $G$ ). It turns out
  that this linear homomorphism is a 1-- cocycle with respect to
  the adjoint action. Moreover, the dual homomorphism $\G^{\ast}
  \wedge \G^{\ast} \rightarrow \G^{\ast}$ satisfies the Jacobi
  identity; i.e. $ \G^{\ast}$ becomes a Lie algebra. Such a pair
  ($\G , \G^{\ast}$) is called a Lie bialgebra \cite{D1}. Each
  Lie bialgebra corresponds to a unique connected, simply
  connected Poisson Lie group. It is easy to show that the pair
  ($\G^{\ast} , \G$) is a Lie bialgebra as soon as ($\G ,
  \G^{\ast}$) is one. The Poisson Lie group ($G^{\ast},
  \pi^{\ast}$) corresponding to ($\G^{\ast} , \G$) will be called
  dual to ($G , \pi$). Thus ( connected, simply connected )
  Poisson Lie groups come in dual pairs.

$\G$ and $ \G^{\ast}$ may be put as Lie subalgebras into the
greater Lie algebra $\widetilde{\G}$ which is called the double Lie
algebra. A vector space  $\widetilde{\G}$ equals $\G \oplus \G^{\ast}$,=20
 with  Lie bracket
\[ [X + \xi , Y + \eta ] =3D
[X,Y] + [\xi , \eta] + ad^{\ast}_{X} \eta - ad^{\ast}_{Y} \xi +
ad^{\ast}_{\xi} Y - ad^{\ast}_{\eta} Y \]

Here  $X, Y \in \G$ , $\xi , \eta \in  \G^{\ast}$  and
$ad^{\ast}$ denotes the coadjoint representations of
$\G$  on  $ \G^{\ast}$ and of   $ \G^{\ast}$ on
$\G =3D( \G^{\ast})^{\ast}$.
 We use [ , ] to denote both the bracket on $\G$ and $
 \G^{\ast}$.

 With respect to $ad$--invariant non-degenerate
canonical bilinear form on $\widetilde{\G}$
\[ (X + \xi , Y + \eta ) =3D \langle X,
\eta \rangle + \langle Y, \xi \rangle \]
$\G$ and $ \G^{\ast}$ form maximal isotropic subspaces of
$\widetilde{\G}$.

The simply connected group $\widetilde{G}$ corresponding to
$\widetilde{\G}$ is the classical Drinfeld double of the Poisson
Lie group ($G , \pi$).

Conversely, any Lie algebra $\widetilde{\G}$ with a
non--degenerate symmetric $ad$--invariant bilinear form and a
pair of  maximal isotropic subalgebras ( a Manin triple )
gives a pair of dual Lie bialgebras by identifying one of the
subalgebras with the dual of the other by means of this bilinear
form.

Let $r=3D \sum_{i} a_{i} \otimes b^{i}$ be an element of $\G
\otimes \G$ ; we say that $r$ satisfies the classical
Yang--Baxter equation if
 \[ [ r_{12}, r_{13} ] + [ r_{12}, r_{23}]
+ [ r_{13}, r_{23} ] =3D 0 .\]
 Here, for instance, $ [ r_{12},
r_{13} ] =3D \sum_{i,j} [a_{i} , a_{j}] \otimes b^{i} \otimes
b^{j}$. A quasitriangular Lie bialgebra is a pair ($\G , r$),
where $\G$ is Lie bialgebra, $r \in \G \otimes \G$, the
coboundary of $r$ is the cobracket $d_{e} \pi : \G \rightarrow \G
\wedge \G$ and $r$ satisfies the classical Yang--Baxter equation.

Let us associate with $r$ a linear operator
\[  r_{+} :    \G^{\ast}  \rightarrow \G, \ \ \ \ \
\xi \mapsto \langle r, \xi \otimes id \rangle .\]
Its adjoint is given by
\[  -r_{-} =3D r_{+}^{\ast}:    \G^{\ast}  \rightarrow \G, \ \ \ \ \
\xi \mapsto \langle r, id \otimes \xi \rangle  =3D
\langle P(r), \xi \otimes id \rangle ,\]
where $P$ is the  permutation operator in $\G \times \G$,
$P(X \otimes Y) =3D Y \otimes X . $

The Lie bracket  [,] in $ \G^{\ast}$ is given by
\[   [ \xi , \eta ] =3D ad_{r_{+}(\xi)}^{\ast} \eta  -
ad_{r_{-}(\eta)}^{\ast} \xi\]

\begin{lemma}[\cite{WX}, \cite{RS} ]

For any quasitriangular Lie bialgebra ($\G , r$), the linear maps
\[   r_{+}, r_{-} : \G^{\ast} \rightarrow \G,  \]
defined above, are both Lie algebra homomorphisms.

\end{lemma}

We now turn our attention to groups.

\begin{defin}[\cite{WX}]

A Poisson Lie group $G$ is called quasitriangular if its
corresponding Lie bialgebra $(\G , \G^{\ast})$ is quasitriangular
and if the Lie algebra homomorphisms $ r_{+}$ and $ r_{-}$ from
$\G^{\ast}$ to $ \G$ lift to Lie group homomorphisms $R_{+}$ and
$R_{-}$ from $G^{\ast}$ to $G$. \end{defin}

It turns out that if $G$ is quasitriangular, the maps $\phi$ and $\psi$
from   $G^{\ast}$ to $G$  are Poisson
morphisms, where $\phi (x) =3D R_{+} (x^{-1}), \  \psi
   (x) =3D R_{-} (x^{-1}) $,
for any $x \in  G^{\ast}$.
For every Poisson Lie group $G$ there
 are naturally defined left and
right ``dressing'' actions of $G$ on $G^{\ast}$ \cite{STS} ,
whose orbits are
exactly the symplectic leaves of $G^{\ast}$. When  $G$ has the zero
Poisson structure, its dual Poisson Lie group is simply $\G^{\ast}$
with the abelian Lie group structure and ordinary Lie--Poisson bracket.
The left and right dressing actions in this case  are simply the left
and right coadjoint actions of $G$ on  $\G^{\ast}$.

Given a Poisson Lie group $(G, \pi)$, we can consider the Lie algebra
homomorphism from $\G$ to the Lie algebra of vector fields on $G^{\ast}$.
To  describe this homomorphism, we pick an element
$ v \in \G =3D (\G^{\ast})^{\ast}. $ It may be identified with an element
$\alpha_{v} \in  T_{e}^{\ast} G^{\ast}$.
 Let $\alpha$ be an extension of  $\alpha_{v} $
  to a right-invariant 1-form
on $G^{\ast}$. Then the vector field $v_{\ast}$ corresponding to $v$
is obtained from $- \alpha$ by means of the
Poisson structure $\pi^{\ast} $.
It turns out that in this way   we get a Lie
 algebra homomorphism from
$\G$  to the Lie algebra of vector fields on $\G^{\ast}$ \cite{W}.
Hence, if all the vector fields $v_{\ast}$ are complete ($G$ is
a complete
Poisson Lie group), we obtain a $G$--action $\lambda$  on
$G^{\ast}$  called the
left dressing action. If in this construction
we replace the right-invariant
1-form by the left-invariant 1-form
on $G^{\ast}$ we obtain the right dressing action $\rho$ of $G$
on $G^{\ast}$.

\begin{defin}[\cite{WX,DP}]

A map $R : S \times S \rightarrow S \times S$,
 where $S$ is any set, is called a
 solution to the set-theoretic quantum
Yang--Baxter equation if
\[   R_{13}  R_{23} R_{12}  =3D R_{12} R_{23} R_{13}, \]
where $R_{ij} : S \times S \times S \rightarrow S \times S \times S$
is $R$ on the $i^{th}$  and $j^{th}$  factors of the
cartesian product and $Id$ on the third one.

\end{defin}

The following Theorem belongs to A. Weinstein and P. Xu  \cite{WX} :

\begin{th} \label{R-mat}
If $G$ is a complete quasitriangular group, then the map
\[ R_{G^{\ast}G^{\ast}} : G^{\ast} \times G^{\ast}  \rightarrow
  G^{\ast} \times G^{\ast} , (u,v) \mapsto
(\lambda_{\psi (v^{-1})} u , \rho_{\phi (u^{-1})} v  ) \]
is a solution to the set-theoretic Yang--Baxter equation.
Moreover, $ R_{G^{\ast}G^{\ast}}$  preserves the Poisson
structure and leaves $\cal O \times O$ invariant for
any symplectic leaf ${\cal O} \subset G=FD^{\ast}$.
\end{th}

By analogy with the case of quantum groups we can get the
braid group action from the solutions to the set-theoretic
Yang-Baxter equation. More precisely,  suppose that
$R : S \times S \rightarrow S \times S$ is a solution to the
set-theoretic Yang-Baxter equation. Let
$\hat{R} =3D  R \circ \sigma$
with $\sigma :  S \times S \rightarrow S \times S$
being the  exchange of components,  and let
$ \hat{R}_{i} (n)$ be the endomorphism of the cartesian power  $S^{n}$
defined by:
\[ \hat{R}_{i} (n) ( (x_{1}, \ldots , x_{n})) =3D
  (x_{1},   \ldots , x_{i-1},
  \hat{R} (x_{i}, x_{i+1}) , x_{i+2} , \ldots , x_{n} ). \]
If $R^{-1}$ exists then by the assignment of  $ \hat{R}_{i} (n)$
to the $i^{th}$ generator $b_{i}$ of the braid group $B_{n}$ we
 obtain an action of $B_{n}$ on $S^{n}$ for each $n$.
This implies that if $G$ is a complete quasitriangular Poisson Lie
group, then $(G^{\ast})^{n} $  admits an action of  $B_{n}$ which
{ \em preserves } the  Poisson structure and which leaves
${\cal O}^{n} $  invariant for
any symplectic leaf ${\cal O} \subset G=FD^{\ast}$.

It is well-known \cite{B},\cite{BZ} that each link can be obtained
 in the standard way from some
braid. In \cite{WX} the following beautiful
fact was proved.

\begin{th}
Suppose that $G$ is a complete quasitriangular Poisson Lie group,
and ${\cal O}_{\mu} \subset G^{\ast}$ is a
symplectic leaf of $G^{\ast} .$
If $A,B \in \amalg_{n} B_{n}$ define
equivalent links, then the fixed point
sets of $A$  and $B$ are diffeomorphic.

\end{th}

The fixed point set from the theorem above is the Weinstein--Xu
invariant of a link. In Sect. 3 we will give topological
interpretation of these invariants in the case of factorizable
Poisson Lie groups.

\pagebreak

 \section{Braids    and representations of link group}

\ \ \ \ \ This section is based on \cite{BAL}. \\
We fix our notation about braid groups first.
 For more details, we refer to \cite{B}.

The Artin's  group $B_{n}$ of braids on $n$ strings
has a standard presentation by generators
$b_{1}, \ldots ,b_{n-1}$
and relations $b_{i} b_{j}   =3D b_{j} b_{i}, \ \  |i-j| > 1$,
and $ b_{i} b_{i+1} b_{i} =3D
b_{i+1} b_{i} b_{i+1}, \ \ \ i=3D1, \ldots ,n-2$.

The group $B_{n}$ can be embedded as a subgroup into $Aut(F_{n})$,
where $F_{n}$   is the free group on
$x_{1}, \ldots , x_{n}$, by setting
\[  b_{i}(x_{j}) =3D  x_{j}, \ \ j \neq i, i+1, \ \ \ \  b_{i}
(x_{i}) =3D x_{i+1}, \ \ \
b_{i} (x_{i+1}) =3D x_{i+1}^{-1}  x_{i}  x_{i+1}     \]

Every oriented link $l$ can be realized as a closed braid $L(b)$
for some braid $b$. If $b$ is an $n$-string braid, the link group
$\pi_{1}(S^{3} - L(b))$ has a representation

\begin{equation} \label{lgr}
\langle x_{1}, \ldots , x_{n} \ | \ b(x_{i}) =3D
x_{i} \ \ \ ( i=3D1, \ldots ,n) \rangle ,
\end{equation}
where we regard $b$ as an element of $Aut(F_{n})$. \\
Let $G$ be an arbitrary group. Then $B_{n}$ acts on the
space of representations of $F_{n}$ into $G$. We denote this action
by $\alpha$. The space of representations equal $G^{n}$, and
a generator $b_{i}$ acts as follows:\begin{equation}  \label{rep}
\alpha (b_{i})  (g_{1}, \ldots , g_{n}) =3D ( g_{1}, \ldots ,
 g_{i-1}, g_{i}^{-1}g_{i+1}g_{i},g_{i},g_{i+2}, \ldots ,g_{n} )
\end{equation}
{}From=20the  representation  (\ref{lgr}) we have
\begin{lemma} \label{l1}
Let a link $l$ be  a closed braid $L(b)$
for some braid $b \in B_{n}$.
Then the space of representations $Hom (\pi_{1}(S^{3} -l ), G )$
can be identified with the fixed point
set of the corresponding action of
$\alpha (b)$ on $G^{n}$.
\end{lemma}

In addition to the above, it is easy to see from the form of action
(\ref{rep}) that if  ${\cal O}_{\mu}$
is an orbit of the conjugation  action,
then braid group $B_{n}$ acts on ${\cal O}_{\mu}^{n}$. Similar to
Lemma \ref{l1}, the fixed point set of the corresponding action of
$\alpha (b)$ on ${\cal O}_{\mu}^{n}$ is the space of representations
of   $\pi_{1}(S^{3} -L(b))$   in $G$ for
which the images of the meridians
belong to the corresponding conjugancy class.


{\bf Example.} Let $G$ be the subgroup of $SL(2,C)$ consisting of all
 upper triangular matrices:
 \[   G =3D \left\{ \left(=20
\begin{array}{cc} t^{-1} & x \\ 0 & t =20
 \end{array}
\right) : x \in C,\ t \in C^{\ast} \right\}. \]
Then  ${\cal O}_{\mu}$ is the subset of $G$ with fixed $t$
and is a vector space.=20
The action  (\ref{rep})  of $B_{n}$ gives us
the Burau representation. Taking the exterior algebra of
 ${\cal O}_{\mu}^{n}$
we obtain the R-matrix and state-models for
the Alexander polynomial of links
\cite{KS}.

\begin{prop}
Let M be the complement of a link $l$ in the 3-sphere.
The following are equivalent.\\
1. There exist a reducible representation of $\pi_{1}(M)$ in $SL(2,C)$
which has non-abelian image and sends the meridians to an elements
 with same eigenvalue $t$.\\
2. $t^{2}$ is a root of the Alexander polynomial of the link $l$.

\end{prop}

Using orbits with different eigenvalues we can obtain the same
result for the multi-variable Alexander polynomials of colored
links \cite{Mur}.

The actions (\ref{rep}) of $B_{n}$ on $G^{n}$ or on ${\cal
O}_{\mu}^{n}$ are the particular cases of the general method of
constructing solutions to the set-theoretic Yang-Baxter equation
from the automorphic sets \cite{BR}.

Recall that a set with product is a pair $(\Delta, \ast)$ where
$\Delta$ is a set and $\ast$ is a map $\Delta \times \Delta
\rightarrow \Delta$. The value of this map for $(a,b)$ will be
denoted by $a \ast b$. A morphism of sets with product $(\Delta,
\ast) \rightarrow (\Delta ', \ast ')$ is a map $\phi : \Delta
\rightarrow \Delta '$ such that $\phi (a \ast b) =3D \phi (a) \ast
' \phi (b)$. For any set with product $(\Delta, \ast)$ and $a \in
\Delta$ the left translation $\lambda_{a}$ is the map
$\lambda_{a} : \Delta \rightarrow \Delta $ defined by
$\lambda_{a} (b) =3D a \ast b$.

\begin{defin}[\cite{BR}]
An automorphic set is a set with product
such that all left translations are automorphisms.
In other words, a  set  with product $(\Delta, \ast)$
is an automorphic set, if it has the following two  properties:
\begin{itemize}
\item  $\forall a,c \in \Delta \ \ \ \ \exists ! \ \ \ \ b \in
            \Delta \ \ \ \ a \ast b =3D c$
\item $\forall a,b,c \in \ \ \ \ \ \ \ \
                 (a \ast b) \ast (a \ast b) =3D a \ast (b \ast c) $
\end{itemize}
\end{defin}

One can find many examples of  automorphic sets in \cite{BR}.
Here I want to note the following one.

{\bf Example. }Let $\Delta$ be 2-dim simply connected manifold
with constant curvature.  Define $a \ast_{\theta} b$ as rotation of $b$
through the angle $\theta$ with center $a$.=20
Then $(\Delta, \ast_{\theta})$
is an automorphic set for all $\theta$.


By elementary calculations one can obtain

\begin{lemma}
If  $(\Delta, \ast)$ is an   automorphic set then the map
\[ R : \Delta \times \Delta \rightarrow \Delta \times \Delta
\ \ \ \ \ \  (a,b) \mapsto (a, a \ast b) \]
 is a solution to the
set-theoretic Yang-Baxter equation.
\end{lemma}

Consequently, for any automorphic set $(\Delta, \ast)$
we have braid group $B_{n}$ action on $\Delta^{n}$.
Recall  that two braids give rise to equivalent links if and only
if they are equivalent under Markov moves.
There are two types of Markov moves:
one is conjugation $A  \rightarrow BAB^{-1}$;
the other is by increasing the number of strings in braid by
a simple twist:
$A \rightarrow A b_{n}^{\pm}$, for $A \in B_{n}$, where $b_{n}$
is $n^{th}$ generator of $B_{n+1}$.
After an elementary calculation we get

\begin{th}
Suppose that  $(\Delta, \ast)$ is an idempotent automorphic set,
i.e. $a \ast a =3D a$  for all $a \in \Delta$. If $A \in B_{n}$
and $B \in B_{m}$ define equivalent links, then the fixed point sets
of $A$ on $\Delta^{n}$ and of $B$ on $\Delta^{m}$ are isomorphic.

\end{th}

{\bf Problem.} {\em Find a topological meaning of  link invariants
from the Theorem above for the idempotent automorphic sets. }
\pagebreak

 \section{Representations into  factorizable Poisson  Lie groups}

Let $r  \in \G \otimes \G$ be the $r-$matrix of a quasitriangular
Lie bialgebra $\G$. The Lie bialgebra $\G$ is called {\em factorizable}
\cite{RS}  if the symmetrization
$I =3D r  + P (r)$      is nondegenerate.
For a factorizable Lie bialgebra the corresponding linear mapping
$j =3D r_{+}  - r_{-}$ is bijective.

Let $G$  be a Poisson Lie group with Lie bialgebra $\G$,
and let $G^{\ast}$ be its simply connected dual.  We can
lift the Lie algebra homomorphisms
$r_{\pm} : \G^{\ast}  \rightarrow \G$
to the  group homomorphisms
 $R_{\pm} : G^{\ast}  \rightarrow G$, and define the map
 $ J : G^{\ast}  \rightarrow G$ by $J(x) =3D R_{+} (x) (R_{-}(x))^{-1} $.
It is not difficult to see that the derivative of $J$ at the identity
element of $G^{\ast}$ is $j$. Note that neither $j$ nor $J$ is a
homomorphism.

The group $G$ is {\em  factorizable} if $J$ is a global diffeomorphism.
In this case for each element $x \in G$ we have a factorization
$x =3D x_{+}  x _{-}^{-1}$, where  $x_{\pm} =3D R_{\pm} (J^{-1} (x))$.
For the factorizable group $G$ Weinstein and Xu \cite{WX}
found an explicit formula for the  solutions to the
set-theoretic Yang-Baxter equation:
\begin{th}

If  $G$ is a  factorizable Poisson
Lie group, under the identification of
$G^{\ast}$ with $G$ via $J$, the  solution to the set-theoretic
Yang--Baxter equation from Theorem \ref{R-mat} takes the form
\begin{equation} \label{9.2}
G \times G \rightarrow G \times G, \ \ \ \ \
(x,y) \mapsto (y_{-} x y_{-}^{-1} ,
 (y_{-} x y_{-}^{-1})_{+}^{-1}  y (y_{-} x y_{-}^{-1})_{+})
\end{equation}

This map is a Poisson diffeomorphism when $G$ is
 equipped with the Poisson structure of $G^{\ast}$.

\end{th}

For the  solution  (\ref{9.2}) to the set-theoretic
Yang--Baxter equation   from the Theorem above
we can construct a braid group action.
The fixed points of this action give us the
link invariant.
\begin{th} \label{m1}

After the following  change of variables
\[  (x^{1} , x^{2}, \ldots , x^{n})
 \rightarrow ( y^{1}, y^{2}, \ldots , y^{n} ) \]
\[    y^{i}   =3D (x^{n})_{-}  \ldots  (x^{i+2})_{-} (x^{i+1})_{-} x^{i}
 (x^{i+1})_{-}^{-1} (x^{i+2})_{-}^{-1}
 \ldots  (x^{n})_{-}^{-1} \]
the Weinstein--Xu action of $B_{n}$ on $G^{n}$ (which is obtained from
(\ref{9.2}) ) is the action induced by  $Aut(F_{n})$  (\ref{rep}).
\end{th}

 It is well-known that the left dressing
action of $G$  on $G^{\ast}$, for factorizable Poisson
Lie group $G$, coincides with the conjugation action
$Ad_{x}$ under the identification of  $G$  with $G^{\ast}$ by $J$.

\begin{th} \label{m2}
Let $G$ be a  factorizable Poisson Lie group and a link
$l$   be a closed braid $L(b)$   for some braid $b \in B_{n}$.
Then the space of  link group representations for which
the images of the meridians belong to the conjugancy
class ${\cal O} \subset G$ is the fixed point set  of a
symplectomorphism ${\cal O}^{n}  \rightarrow  {\cal O}^{n}$.
\end{th}

{\bf Example.}  It is well-known \cite{WL,M} that $SL(2,C)$
is a factorizable Poisson Lie group with dual
 $(SL(2,C))^{\ast} =3D SU(2,C) \times  SB(2,C)^{\sigma}$,
where  $SB(2,C)$  is the ``book group'' consisting of all
upper triangular matrices with positive diagonal and
determinant 1.
The factorization $x=3D x_{+} x_{-}^{-1}$ for $x \in SL(2,C)$
is exactly the Gram--Shmidt process with
$x_{+} \in  SU(2,C)$  and  $x_{-} \in  SB(2,C)$ .
 Theorems \ref{m1}, \ref{m2} imply that for a link
$l =3D L(b)$, $b \in B_{n}$  the space of representations
$Hom(\pi_{1} (S^{3} - L(b)) , SL(2, C) )$ is the fixed set
of the  {\em Poisson diffeomorphism}
\[  b : SL(2,C)^{n}  \rightarrow  SL(2,C)^{n}  \]
and  the space of the link groups representations for which
the images of the meridians belong to the conjugancy
class ${\cal O} \subset SL(2,C)$ is the fixed point set  of the
symplectomorphism ${\cal O}^{n}  \rightarrow  {\cal O}^{n}$.

It will be very interesting to investigate the connection
 of this Poisson structure on $SL(2,C)$ with the geometrical
approach to $Hom(\pi_{1} (S^{3} - L(b)) , SL(2, C) )$  from \cite{CC}
and with the group actions on trees \cite{Sh},\cite{Mor}.

Let $G=3DSL(2,C)$, and let $H,\ X_{+}, \ X_{-}$
be the standard generators
of its Lie algebra $sl(2,C)$. Then the quasitriangular
$r-$matrix $r =3D X_{+} \otimes X_{-} + \frac{1}{4} H \otimes H
 \in  sl(2,C) \otimes sl(2,C)$ defines a
 complex Poisson Lie structure on
$SL(2,C)$. The dual of $SL(2,C)$ is a subgroup
$B_{+} \times B_{-}$
 of $SL(2,C) \times SL(2,C)$ which
 consists of all the elements of the form

\[ \left\{  \left( \begin{array}{cc} \lambda & f \\ 0 & \lambda^{-1}
 \end{array} \right) ,
 \left( \begin{array}{cc} \lambda^{-1} & 0 \\ -g & \lambda
 \end{array}  \right)
 :  \forall f,g \in G , \lambda \in C^{\ast}  \right\}        \]

The maps $ R_{\pm}: B_{+} \times B_{-}   \rightarrow SL(2,C)$ are,
respectively, the
 projection to the first and second factors.
Let $K$ be the image of the map
$J: B_{+} \times B_{-}   \rightarrow SL(2.C),
J(x) =3D R_{+}(x) R_{-}(x)^{-1}$.
Then $K$ is an open dense subset of $SL(2,C)$.
The decomposition for $x \in K$ into
$x_{+} x_{-}^{-1}$ is not unique.
In fact there are two ways of decomposing $x$ and  the
resulting $x_{+}$ and  $x_{-}$ differ by sign.
So the map (\ref{9.2}) and the
change of variables from Theorem \ref{m1} are still  well-defined,
whenever their right-hand sides are defined.
Therefore, they satisfy the quantum Yang-Baxter equation, but
we do not know the appropriate domain for these maps.




\pagebreak

\end{document}